\documentclass{article}
\usepackage{amssymb}
\usepackage{amsmath}
\usepackage{amsfonts}
\usepackage{version}
\usepackage{appendix}
\usepackage{portland}
\usepackage{lscape}
\usepackage{setspace}

\newtheorem{theorem}{Theorem}

\newtheorem{definition}{Definition}

\newtheorem{proposition}{Proposition}
\newtheorem{remark}{Remark}

\newenvironment{proof}[1][Proof]{\noindent \textbf{#1.} }{\  \rule{0.5em}{0.5em}}
\input{tcilatex}
\begin{document}

\title{A univariate time varying analysis of periodic ARMA processes }
\author{M. Karanasos$^{\dagger }$, A. G. Paraskevopoulos$^{\ddagger }$ and
S. Dafnos \\
$^{\dagger }$\textit{Brunel University, London, UK}\\
$^{\ddagger }$University of Patras, Patra, Greece}
\date{This draft: March 18th 2014\\
}
\maketitle

\begin{abstract}
The standard approach for studying the periodic ARMA model with coefficients
that vary over the seasons is to express it in a vector form. In this paper
we introduce an alternative method which \ views the periodic formulation as
a time varying univariate process and obviates the need for vector analysis.
The specification, interpretation, and solution of a periodic ARMA process
enable us to formulate a forecasting method which avoids recursion and
allows us to obtain analytic expressions of the optimal predictors. Our
results on periodic models are general, analogous to those for stationary
specifications, and place the former on the same computational basis as the
latter.


\textbf{JEL Classifcation}: C22, C53, C58.

\vspace{1.5in}

{\footnotesize We gratefully} {\footnotesize acknowledge the helpful
conversations we had with L. Giraitis, G. Kapetanios and A. Magdalinos in
the preparation of the paper. We would also like to thank R. Baillie, L.
Bauwens, M. Brennan, D. van Dijk, W. Distaso, C. Francq, P. Fryzlewicz, C.
Gourieroux, E. Guerre, M. Guidolin, A. Harvey, C. Hommes, S. Leybourne, P.
Minford, A. Monfort, C. Robotti, W. Semmler, R. Smith, T. Ter\"{a}svirta, P.
Zaffaroni, and J-M. Zakoian for suggestions and comments on a closely
related work (see Paraskevopoulos, Karanasos and Dafnos 2013) which greatly
improved many aspects of the current paper as well. We are grateful to
seminar participants at CREST, Erasmus University, London School of
Economics, Queen Mary University of London, University of Essex, Birkbeck
College University of London, University of Nottingham, Cardiff University,
University of Manchester, Athens University of Economics and Business, and
University of Piraeus. We have also benefited from the comments given by
participants (on the closely related work) at the 3rd Humboldt-Copenhagen
Conference on Financial Econometrics (Humboldt University, Berlin, March
2013), the SNDE 21st Annual Symposium (University of Milan-Bicocca, March
2013), the 8th and 9th BMRC-QASS conferences on Macro and Financial
Economics (Brunel University, London, May 2013), the 7th CFE Conference
(Senate House, University of London, December 2013), and the 1st RASTANEWS
Conference (University of Milan-Bicocca, January 2014).}

$^{\dagger }${\footnotesize Address for correspondence: Menelaos Karanasos,
Economics and Finance, Brunel University, West London, UB3 3PH, UK; email:
menelaos.karanasos@brunel.ac.uk, tel: +44(0)1895265284, fax: +44
(0)1895269770.}
\end{abstract}

\newpage

\doublespacing

\section{INTRODUCTION}

Many natural and biological phenomena are dominated by the existence of
periodic regularities, which in economics are due to seasonality.\footnote{%
The property of periodicity is important in many fields. \ \ For the
applications of periodic time series models in climatology, hydrology and
electrical engineering see the references cited in: Lund and Basawa (2000),
Basawa and Lund (2001) and Shao (2008). \ The periodic models were applied
in economics by Parzen and Pagano (1979). \ They became popular in the then
emerging subject of macroeconometrics in the mid to late 1980's with Miron
(1986), Ghysels (1988), Osborn (1988, 1990), and in joint work with their
co-authors. For modern treatments and overviews see Franses (1996b), Ghysels
and Osborn (2001), Franses and Paap (2004), and Hurd and Miamee (2007). \
All the developments we have referred to above share a common theme: they
all point to the importance of periodicities in the analysis of time series,
which are subject to seasonal fluctuations.} \ To express periodicities,
Gladyshev (1961) introduced a mathematical model that still constitutes the
core of the prevailing approach to the analysis of seasonal time series. \
Employing Gladyshev's results, the bulk of the literature transforms the
problem of investigating a periodic univariate series to the corresponding
problem for stationary vector series. In this paper we propose a theory by
which we investigate the time series properties of periodic\ schemes. \ We
consider them as univariate time varying frameworks (instead of time
invariant multivariate ones), that is we regard them as stochastic
difference equations with time dependent (albeit periodically varying)
parameters.

Subsequent literature has embodied Gladyshev's approach in the mainstream
theory of time series. \ Jones and Brelsford (1967) and Troutman (1979)
modeled periodic time series as autoregressive processes. \ Gladyshev's
scheme was incorporated into an autoregressive moving average framework by
Cleveland and Tiao (1979) and developed further by Tiao and Grupe (1980)
(see also Vecchia 1985; Osborn, 1991). \ The outcome of this research
program is the periodic autoregressive moving average, or PARMA model. \ In
the present paper we introduce a new method for the study of periodic models
with coefficients that vary over the seasons, as an alternative to the
standard approach of expressing them in a vector form. \ Viewing the
periodic formulation as a time varying (TV) univariate process obviates the
need for vector analysis. \ As we explain below, our results on PARMA models
are general, analogous to those for stationary ARMA specifications, and
place the former on the same computational basis as the latter.

The standard modeling of PARMA formulations, as expressed, say, in the
influential papers by Tiao and Grupe (1980) and Osborn (1991), treats them
as nonperiodic vector models in order to study their properties. \ In other
words, they examine the periodic specification by converting it into a
vector ARMA (VARMA) process with constant coefficients. \ However, Lund,
Shao, and Basawa (2006) call attention to the fact that the time invariant
vector form even of a periodic autoregressive model of order one for daily
data will contain 365 variables, and this is a handicap, especially for
forecasting.

Except for some notable exceptions, (see for example, Vecchia, 1985;
Franses, 1994, 1996a; Lund and Basawa, 2000; Franses and Paap, 2005), the
time series properties of periodic processes have not been fully
investigated.\footnote{%
Difficulties in testing for unit roots in seasonal models have been
investigated by Franses (1991, 1994), Taylor (2002) and del Barrio Castro
and Osborn (2008; see the references therein for this stream of important
research). Basawa and Lund (2001), Shao (2005), and Tesfaye, Anderson, and
Meerschaert (2011) discuss parameter estimation and asymptotic properties of
PARMA specifications.} \ Lund and Basawa (2000) propose a recursive scheme
for computing one-step ahead predictors for such processes, and construct
multi-step-ahead forecasts recursively from the one-step ahead predictions.
\ Anderson, Meerschaert, and Zhang (2013) develop a recursive forecasting
algorithm for periodic processes. \ But, as pointed out by Lund and Basawa
(2000), despite their applicability, prediction for PARMA models remains
relatively unexplored, compared to their stationary counterparts. \ Although
they consider recursive computation of linear predictors and their mean
squared errors, as we will show below, our explicit solution of the PARMA
formulation frees us from the bounds of recursion and enables us to derive
formulas that facilitate the analytic calculation of the multi-step-ahead
forecasts.

We put forward the solution to periodic schemes, which is based on the
representation of the PARMA model as an infinite system of linear equations;
its coefficient matrix is row-finite, that is, an infinite matrix whose rows
comprise a finite number of non-zero entries. \ This solution is derived
from a general method for solving infinite linear systems in row-finite
form, developed recently by Paraskevopoulos (2012). \ It is an efficient
systematic procedure that generalizes the Gauss-Jordan elimination;
implemented under a rightmost pivoting, it solves the infinite systems where
the standard Gauss-Jordan elimination fails.

Once we have expressed the PARMA model as an infinite linear system, we only
need the infinite Gaussian part of the Gauss-Jordan algorithm. \ This is due
to the fact that the row-finite coefficient matrix has the additional
property of being in row-echelon form. \ The application of the Gaussian
algorithm to time varying linear difference equations leads to solutions
expressed in terms of a single Hessenbergian, which in our case is the
determinant of a lower Hessenberg matrix \ \ The solution derived by the
approach described above is decomposable into two parts: the homogeneous and
particular solutions which are also expressed in terms of a single
Hessenbergian. \ For the periodic processes that we study in the current
paper, the coefficients in these solutions are expressed as generalized
lower continuant matrices, which are special forms of Hessenbergians. \ This
allows us to provide a characterization of PARMA models by deriving, first,
multistep ahead forecasts, the associated forecast error, and the mean
square error, and second, the first two unconditional moments of the process
and its covariance structure. \ Our predictions can be employed to develop
an efficient algorithm for the PARMA likelihood of Gaussian series, as in
Lund and Basawa (2000). Equally important we relax the assumption of
homoscedasticity (see also, among others, Paraskevopoulos, Karanasos, and
Dafnos, 2013, and Karanasos, Paraskevopoulos, Menla Ali, Karoglou, and
Yfanti, 2014), which is likely to be violated in practice, and allow $%
\varepsilon _{t}$\ to follow, for example, a periodical GARCH type of
process (see, Bollerslev and Ghysels, 1996).

The paper is organized as follows. \ First we introduce suitable seasonal
notation in Section $2.1$ and then in Section $2.2$ we state the stochastic
periodic difference equation, which is our main object of inquiry. In
Section $3.1$ we represent this equation as an infinite linear system and
concentrate on the associated coefficient matrix. \ After appropriate
transformations of this matrix we end up with the fundamental solution
matrices, which are band matrices with a superdiagonal of non-zero entries;
they belong to the class of lower Hessenberg matrices. In Section $3.2$ we
employ their determinants, called the Hessenbergians, in order to express
the general solution of the periodic model as the sum of two parts: the
homogeneous and particular solutions. \ In Section $4$, we derive the
fundamental properties of the PARMA model. \ For example, simplified
closed-form expressions of the multi-step forecast error variances are
obtained. These formulas allow a fast computation of the multi-step-ahead
predictors. Section $5$ concludes, offers suggestions for future research
and reflects on the significance and appropriate approach to studying time
series data subject to periodicities. The proof of the general solution
theorem is in the Appendix A. \ Appendix B helps us to understand the
difference between a stationary treatment of periodic processes and the time
varying approach of the present paper

\section{PROLEGOMENA\label{SecTVARMA}}

In the current section we introduce suitable seasonal notation, and then we
describe the problem we study.

\subsection{Seasonal Notation}

Throughout the paper we adhere to the following conventions: ($\mathbb{Z}%
^{+} $) $\mathbb{Z},$ and ($\mathbb{R}^{+}$) $\mathbb{R}$ stand for the sets
of (positive) integers, and (positive) real numbers, respectively. To
simplify our exposition we introduce the following notation: ($t,T$)$\in $ $%
\mathbb{Z}\times \mathbb{Z}$, ($n,l$) $\in $ $\mathbb{Z}^{+}\times \mathbb{Z}%
^{+}$.

Next consider a time series subject to periodic fluctuations. \ The periodic
notation $y_{Tl+s}$ denotes the series during the $s$th season, $s=1,\ldots
,l$, where $l$ denotes the number of seasons (e.g. quarters in a year: $l=4$%
); so, $l$ is the length of the period. \ $T$ is the number of periods (e.g.
years in our example); that is, $T$ $=0,\ldots ,n$. \ For clarity of
notation, we denote the sum $Tl+s$ with the symbol $t_{s}$; accordingly, we
indicate a variable that stands for a periodic series as $y_{t_{s}}$. \ The
present time is represented by $t$, and $nl$ is the number of seasons such
that at time $\tau _{n}=t-nl$ information is given.\footnote{%
We assume that information is given at time $\tau _{n}=t-nl$ for ease of
exposition. It can, of course, be given at any time $\tau _{n}-s$.}

\subsection{PARMA Model\label{SubsecNotDef}}

Next we give the main definition that we will use in the rest of the paper.

\begin{definition}
\label{DefTVARMA}We can write a periodic $AR$ model of order $p$ with $l$
seasons, $PAR(p;l)$, as%
\begin{equation}
y_{t_{s}}=\phi _{0,s}+\sum_{m=1}^{p}\phi _{m,s}y_{t_{s}-m}+\varepsilon
_{t_{s}},  \label{PAR(P)}
\end{equation}%
which can be written in a more efficient way using the backshift operator, $%
B $ as%
\begin{equation*}
\Phi _{t_{s}}(B)y_{t_{s}}=\phi _{0,s}+\varepsilon _{t_{s}},
\end{equation*}%
where time $t_{s}=Tl+s$ is at the $sth$ season, and $\phi _{m,s}$ are the
periodically (or seasonally) varying autoregressive coefficients. \ For
example, if $s=l$, that is, we are at the last season (which for quarterly
data is the fourth), then the periodically varying coefficients are $\phi
_{m,l}$ ($\phi _{m,4}$); whereas, if $s=1$ we are at the first season and
thus the periodically varying coefficients are $\phi _{m,1}$. A periodically
varying drift is denoted by $\phi _{0,s}$; $\{ \varepsilon _{t_{s}},t_{s}\in 
\mathbb{Z}
\}$ is a sequence of zero mean serially uncorrelated random variables
defined on $L_{2}(\Omega ,\tciFourier _{t},P)$\footnote{%
The triple $(\Omega ,\{ \tciFourier _{t},t\in 
\mathbb{Z}
\},P)$\ denotes a complete probability space with a filtration, $\{
\tciFourier _{t}\}$,\ which is a non-decreasing sequence of $\sigma $-fields 
$\tciFourier _{t-1}\subseteq \tciFourier _{t}\subseteq \tciFourier $, $t\in 
\mathbb{Z}
$. The space of $P$-equivalence classes of finite complex random variables
with finite $p$-order is indicated by $L_{p}$. Finally, $H=L_{2}(\Omega
,\tciFourier _{t},P)$\ stands for a Hilbert space of random variables with
finite first and second moments.} with $\mathbb{E}[\varepsilon
_{t_{s}}\left
\vert \tciFourier _{t_{s}-1}\right. ]=0$ a.s., and finite
variance: $0<M_{l}<\sigma _{t_{s}}^{2}<M<\infty $, $\forall $ $t_{s}$, for
some $M_{l}\in \mathbb{R}^{+}$ and $M\in \mathbb{R}^{+}$. 
$\Phi _{t_{s}}(B)$ is a $p$-order polynomial of the backshift
operator $B$ with periodical coefficients $\phi _{m,s}$. The above process
nests the AR($p$) model as a special case, i.e , it reduces to the AR($p$)
process if we assume that the drift and all the AR parameters are constant,
that is: $\phi _{m,s}=\varphi _{m}$, $m=0,\ldots ,p$, for all $t$. To obtain
a PARMA($p,q;l$) model we replace $\varepsilon _{t_{s}}$ by $%
u_{t_{s}}=\Theta _{t_{s}}(B)\varepsilon _{t}=\sum_{j=1}^{q}\theta
_{j,s}\varepsilon _{t_{s}-j}$.%
\end{definition}

\section{GENERAL SOLUTION IN TERMS OF HESSENBERGIANS}

In the current section we put forward a framework for examining periodic
time series models, like eq. (\ref{PAR(P)}), based on a workable closed form
solution of higher order stochastic time varying difference equations. We
introduce a method for finding the $p$ linearly independent solutions that
we need in order to obtain the general solution of the PAR($p;l$) process,
the so called fundamental solutions.

\subsection{Fundamental Solution Matrices}

Gladyshev (1961) introduced a category of non-stationary time series, called
periodically correlated; such series exhibit periodic means and covariances.
Gladyshev bypassed the non-stationarity of a univariate periodic series with 
$l$ periods, by representing it as an $l$-dimensional (i.e. multivariate)
stationary vector series. Building on the seminal Gladyshev paper, Tiao and
Grupe (1980), Osborn (1991), and the bulk of the subsequent literature have
modeled periodic autoregressions as VAR models with constant coefficients.

Departing from this tradition, we face the non-stationarity of periodic
processes head on with a time varying treatment by staying within the
univariate framework. We express the periodic difference equation (\ref%
{PAR(P)}) as an infinite system and provide its explicit solution.

A main advantage associated with our time varying analysis of periodic
processes is that we avoid a major drawback of the standard stationary
multivariate approach, namely, that it might require the consideration of a
large number of variables. This weakness becomes especially acute in the
examination of high frequency data.

We begin by expressing the PAR($p;l$) model as a time varying AR model. That
is, 
\begin{equation}
\Phi _{t}(B)y_{t}=\phi _{0}(t)+\varepsilon _{t}\text{, }  \label{TVAR(P)}
\end{equation}%
with $\Phi _{t}(B)=1-\sum_{m=1}^{p}\phi _{m}(t)B^{m}$ where $\phi
_{m}(t)=\phi _{m}(t-nl)$, $m=1,\ldots ,p$ are the periodically (or
seasonally) varying autoregressive coefficients: $\phi _{m,s}\triangleq \phi
_{m}(Tl+s)$. Similarly, for the PARMA process we can replace $\varepsilon
_{t}$ with $u_{t}=\Theta _{t}(B)\varepsilon _{t}$ and $\Theta
_{t}(B)=\sum_{j=1}^{q}\theta _{j}(t)B^{j}$ where $\theta _{j}(t)=\theta
_{j}(t-nl)$, $j=1,\ldots ,q$, are the periodically varying moving average
coefficients: $\theta _{j,s}\triangleq \theta _{j}(Tl+s)$.

Equation (\ref{TVAR(P)}) is written as 
\begin{equation}
\sum_{m=1}^{p}\phi _{m}(t)y_{t-m}-y_{t}=-[\phi _{0}(t)+\varepsilon _{t}],
\label{Difference(P)}
\end{equation}%
and takes the infinite band system form 
\begin{equation}
\mathbf{\Phi \cdot y}=-\mathbf{\phi }-\mathbf{\varepsilon ,}
\label{PHIMatrix}
\end{equation}%
(matrices and vectors are denoted by upper and lower case boldface symbols,
respectively) where

\begin{equation*}
\mathbf{\Phi =}\left( 
\begin{array}{ccccccccc}
\phi _{p}(\tau _{n}+1) & \phi _{p-1}(\tau _{n}+1) & ... & \phi _{1}(\tau
_{n}+1) & -1 & 0 & 0 & 0 & ... \\ 
0 & \phi _{p}(\tau _{n}+2) & ... & \phi _{2}(\tau _{n}+2) & \phi _{1}(\tau
_{n}+2) & -1 & 0 & 0 & ... \\ 
0 & 0 & ... & \phi _{3}(\tau _{n}+3) & \phi _{2}(\tau _{n}+3) & \phi
_{1}(\tau _{n}+3) & -1 & 0 & ... \\ 
\vdots & \vdots & \vdots \vdots \vdots & \vdots & \vdots & \vdots & \vdots & 
\vdots & \vdots \vdots \vdots%
\end{array}%
\right) ,
\end{equation*}%
with%
\begin{equation*}
\mathbf{y}\mathbf{=}\mathbf{(}y_{\tau _{n}+1-p},\text{ }y_{\tau _{n}+2-p},%
\text{ }\ldots ,\text{ }y_{\tau _{n}},\text{ }y_{\tau _{n}+1},\text{ }%
y_{\tau _{n}+2},\text{ }y_{\tau _{n}+3},\text{ }y_{\tau _{n}+4},\text{ }%
\ldots \mathbf{)}^{\prime },
\end{equation*}%
and%
\begin{equation*}
\mathbf{\phi }\mathbb{=}\left( 
\begin{array}{l}
\phi _{0}(\tau _{n}+1) \\ 
\phi _{0}(\tau _{n}+2) \\ 
\phi _{0}(\tau _{n}+3) \\ 
\multicolumn{1}{c}{\vdots}%
\end{array}%
\right) \text{, }\mathbf{\varepsilon =}\left( 
\begin{array}{l}
\varepsilon _{\tau _{n}+1} \\ 
\varepsilon _{\tau _{n}+2} \\ 
\varepsilon _{_{\tau _{n}}+3} \\ 
\multicolumn{1}{c}{\vdots}%
\end{array}%
\right)
\end{equation*}%
(recall that $\tau _{n}=t-nl$). The elements of the matrices $\mathbf{\phi }$%
, and $\mathbf{\Phi }$ are the values that their respective coefficients
take in successive time periods. The equivalence of (\ref{Difference(P)})
and (\ref{PHIMatrix}) follows from the fact that the $i$th equation in (\ref%
{PHIMatrix}), as a result of the multiplication of the $i$th row of $\mathbf{%
\Phi }$ by the column of $y${\footnotesize s} equated to $-[\phi _{0}(\tau
_{n}+i)+\varepsilon _{_{\tau _{n}+i}}]$, is equivalent to eq. (\ref%
{Difference(P)}), as of time $\tau _{n}+i$. The $\mathbf{\Phi }$ matrix in
eq. (\ref{PHIMatrix}) can be partitioned as 
\begin{equation*}
\mathbf{\Phi =}\left( 
\begin{tabular}{l|l}
$\mathbf{P}$ & $\mathbf{C}$%
\end{tabular}%
\right) ,
\end{equation*}%
where%
\begin{equation}
\mathbf{P=}\left( 
\begin{array}{cccc}
\phi _{p}(\tau _{n}+1) & \phi _{p-1}(\tau _{n}+1) & ... & \phi _{1}(\tau
_{n}+1) \\ 
0 & \phi _{p}(\tau _{n}+2) & ... & \phi _{2}(\tau _{n}+2) \\ 
0 & 0 & ... & \phi _{3}(\tau _{n}+3) \\ 
\vdots & \vdots & \vdots \vdots \vdots & \vdots%
\end{array}%
\right) ,\text{ }\mathbf{C=}\left( 
\begin{array}{ccccc}
-1 & 0 & 0 & 0 & ... \\ 
\phi _{1}(\tau _{n}+2) & -1 & 0 & 0 & ... \\ 
\phi _{2}(\tau _{n}+3) & \!\!\phi _{1}(\tau _{n}+3) & -1 & 0 & ... \\ 
\vdots & \vdots & \vdots & \vdots & \vdots \vdots \vdots%
\end{array}%
\right) .  \tag{4a}
\end{equation}%
The matrix $\mathbf{P}$ consists of the first $p$ columns of $\mathbf{\Phi }$
and the $j$th column of $\mathbf{C}$, $j=1,2,\ldots ,$ is the ($p+j$)th
column of $\mathbf{\Phi }$. We will denote the $p$th column of the $nl\times
p$ top submatrix of the matrix $\mathbf{P}$ by $\mathbf{\phi }_{t,nl}$:%
\begin{equation*}
(\mathbf{\phi }_{t,nl})^{\prime }=\left( 
\begin{array}{lllllll}
{\small \phi }_{1}{\small (\tau }_{n}{\small +1),} & {\small \phi }_{2}%
{\small (\tau }_{n}{\small +2),} & \ldots & ,{\small \phi }_{p}{\small (\tau 
}_{n}{\small +p),} & 0, & \ldots & ,0%
\end{array}%
\right)
\end{equation*}%
(assuming without loss of generality that $p<nl$).

The $nl\times (nl-1)$ top submatrix of matrix $\mathbf{C}$ is called the
core solution matrix and is denoted as 
\begin{equation}
\mathbf{C}_{t,nl}=\left( 
\begin{array}{cccccccc}
-1 &  &  &  &  &  &  &  \\ 
\phi _{1}(\tau _{n}+2) & -1 &  &  &  &  &  &  \\ 
\phi _{2}(\tau _{n}+3) & \phi _{1}(\tau _{n}+3) & \ddots &  &  &  &  &  \\ 
\vdots & \vdots & \ddots & \ddots &  &  &  &  \\ 
\phi _{p}(\tau _{n}+p+1) & \phi _{p-1}(\tau _{n}+p+1) & \ddots & \ddots & 
\ddots &  &  &  \\ 
& \phi _{p}(\tau _{n}+p+2) & \ddots & \ddots & \ddots & \ddots &  &  \\ 
&  & \ddots & \ddots & \ddots & \ddots & \ddots &  \\ 
&  &  & \phi _{p}(t-1) & \phi _{p-1}(t-1) & \cdots & \phi _{1}(t-1) & -1 \\ 
&  &  &  & \phi _{p}(t) & \cdots & \phi _{2}(t) & \phi _{1}(t)%
\end{array}%
\right) ,  \label{CoreMatrix}
\end{equation}%
(here and in what follows empty spaces in a matrix have to be replaced by
zeros). The fundamental solution matrix is obtained from the core solution
matrix $\mathbf{C}_{t,nl}$, augmented on the left by the $\mathbf{\phi }%
_{t,nl}$ column:

\begin{eqnarray}
\mathbf{\Phi }_{t,nl} &=&\left( 
\begin{tabular}{l|l}
$\mathbf{\phi }_{t,nl}$ & $\mathbf{C}_{t,nl}$%
\end{tabular}%
\right) =  \notag \\
&&\left( 
\begin{array}{cccccccc}
\phi _{1}(\tau _{n}+1) & -1 &  &  &  &  &  &  \\ 
\phi _{2}(\tau _{n}+2) & \phi _{1}(\tau _{n}+2) & \ddots &  &  &  &  &  \\ 
\vdots & \vdots & \ddots & \ddots &  &  &  &  \\ 
\phi _{p}(\tau _{n}+p) & \phi _{p-1}(\tau _{n}+p) & \ddots & \ddots & \ddots
&  &  &  \\ 
& \phi _{p}(\tau _{n}+p+1) & \ddots & \ddots & \ddots & \ddots &  &  \\ 
&  & \ddots & \ddots & \ddots & \ddots & \ddots &  \\ 
&  &  & \phi _{p}(t-1) & \phi _{p-1}(t-1) & \cdots & \phi _{1}(t-1) & -1 \\ 
&  &  &  & \phi _{p}(t) & \cdots & \phi _{2}(t) & \phi _{1}(t)%
\end{array}%
\right) .  \notag \\
&&  \label{FAIMAR(p)X}
\end{eqnarray}

The entries of the above $nl\times nl$ matrix are given by:

\begin{equation*}
\left \{ 
\begin{array}{cccccc}
-1 & \text{if} &  & i=j-1, & \text{and} & 2\leq j\leq nl, \\ 
\phi _{1+m}(t-nl+i) & \text{if} & \text{ }0\leq m\leq p-1, & i=j+m, & \text{%
and} & 1\leq j\leq nl-m, \\ 
0 &  & \text{otherwise.} &  &  & 
\end{array}%
\right.
\end{equation*}

The solution matrix, $\mathbf{\Phi }_{t,nl}$ for $p\leq nl$, is a $(p+1)$%
-diagonal matrix of order $nl$, that is a matrix that possesses $p+1$\
diagonals with nonzero entries. Apart from the main diagonal, the
superdiagonal, and the subdiagonal, it also possesses $p-2$\ nonzero time
varying lower diagonals. Therefore, we call it a generalized lower
continuant matrix of degree $p+1$. When $p=2$\ then all lower diagonals are
zero and $\mathbf{\Phi }_{t,nl}$\ becomes a continuant or a tridiagonal
matrix (see Karanasos, Paraskevopoulos, Menla Ali, Karoglou, and Yfanti,
2014).

Next we introduce the bivariate function $\xi :\mathbb{Z}\times \mathbb{Z}%
^{+}\longmapsto \mathbb{R}$ by%
\begin{equation}
\xi _{t,nl}=\text{det}(\mathbf{\Phi }_{t,nl})  \label{KSIAR(p)}
\end{equation}%
(for square matrices $\mathbf{X}=[x_{ij}]_{i,j=1,\ldots ,l}\in \mathbb{R}%
^{l\times l}$ using standard notation, det$(\mathbf{X})$ or $\left\vert 
\mathbf{X}\right\vert $ denotes the determinant of matrix $\mathbf{X}$)
coupled with the initial values $\xi _{t,0}=1$, and $\xi _{t,-m}=0$ for $%
m=1,\ldots ,p-1$. In other words, $\xi _{t,nl}$ is the determinant of an $%
nl\times nl$ matrix; each nonzero diagonal of this matrix, below the
superdiagonal, consists of the periodical coefficients $\phi _{r}(\mathfrak{%
\cdot })$, $r=1,\ldots ,\min (nl,p)$ from $t-nl+r$ to $t$. In other words, $%
\xi _{t,nl}$ is an\ $nl$-order generalized lower continuant determinant of
degree $p+1$. Note that $\xi _{t,nl-r}=$det$(\mathbf{\Phi }_{t,nl-r})$ for $%
r<nl$, where $\mathbf{\Phi }_{t,nl-r}$ is equal to the matrix $\mathbf{\Phi }%
_{t,nl}$ without its first $r$ rows and columns.

The solution matrices, being band matrices with a superdiagonal of non-zero
elements, are special cases of lower Hessenberg matrices, the determinants
of which are called Hessenbergians.

Alternatively $\mathbf{\Phi }_{t,nl}$ can be written as

\begin{equation}
\mathbf{\Phi }_{t,nl}=\left( 
\begin{array}{lllll}
\mathbf{\Phi }_{\tau _{n-1},l} & \overline{\mathbf{0}} &  &  &  \\ 
\widetilde{\mathbf{0}}_{\tau _{n-2}} & \mathbf{\Phi }_{\tau _{n-2},l} & 
\overline{\mathbf{0}} &  &  \\ 
& \ddots & \ddots & \ddots &  \\ 
&  & \widetilde{\mathbf{0}}_{\tau _{1}} & \mathbf{\Phi }_{\tau _{1},l} & 
\overline{\mathbf{0}} \\ 
&  &  & \widetilde{\mathbf{0}}_{t} & \mathbf{\Phi }_{t,l}%
\end{array}%
\right) ,  \label{BlockToeplitz}
\end{equation}%
where $\overline{\mathbf{0}}$ is an $l\times l$ matrix of zeros except for $%
-1$ in its ($l,1$) entry; $\widetilde{\mathbf{0}}_{t}$ is an $l\times l$
matrix of zeros except $\phi _{m}(t-l+i)$, in its ($i,l-m+i+1$) entry, $%
m=i+1,\ldots ,\min (p,l+2)$. Since $\phi _{t_{T},l}=\phi _{t,l}$: $%
\widetilde{\mathbf{0}}_{\tau _{T}}=\widetilde{\mathbf{0}}_{t}$, and $\mathbf{%
\Phi }_{\tau _{T},l}=\mathbf{\Phi }_{t,l}$, we have%
\begin{equation*}
\mathbf{\Phi }_{t,nl}=\left( 
\begin{array}{lllll}
\mathbf{\Phi }_{t,l} & \overline{\mathbf{0}} &  &  &  \\ 
\widetilde{\mathbf{0}}_{t} & \mathbf{\Phi }_{t,l} & \overline{\mathbf{0}} & 
&  \\ 
& \ddots & \ddots & \ddots &  \\ 
&  & \widetilde{\mathbf{0}}_{t} & \mathbf{\Phi }_{t,l} & \overline{\mathbf{0}%
} \\ 
&  &  & \widetilde{\mathbf{0}}_{t} & \mathbf{\Phi }_{t,l}%
\end{array}%
\right) .
\end{equation*}%
The above matrix\textbf{\ }is a block Toeplitz matrix of bandwidth $3$. $%
\mathbf{\Phi }_{t,l}$ is the $\mathbf{\Phi }_{t,nl}$ matrix defined in eq. (%
\ref{FAIMAR(p)X}) when $n=1$: 
\begin{equation}
\mathbf{\Phi }_{t,l}=\left( 
\begin{array}{cccccccc}
\phi _{1}(\tau _{1}+1) & -1 &  &  &  &  &  &  \\ 
\phi _{2}(\tau _{1}+2) & \phi _{1}(\tau _{1}+2) & \ddots &  &  &  &  &  \\ 
\vdots & \vdots & \ddots & \ddots &  &  &  &  \\ 
\phi _{p}(\tau _{1}+p) & \phi _{p-1}(\tau _{1}+p) & \ddots & \ddots & \ddots
&  &  &  \\ 
& \phi _{p}(\tau _{1}+p+1) & \ddots & \ddots & \ddots & \ddots &  &  \\ 
&  & \ddots & \ddots & \ddots & \ddots & \ddots &  \\ 
&  &  & \phi _{p}(t-1) & \phi _{p-1}(t-1) & \cdots & \phi _{1}(t-1) & -1 \\ 
&  &  &  & \phi _{p}(t) & \cdots & \phi _{2}(t) & \phi _{1}(t)%
\end{array}%
\right) .  \label{FAIM1}
\end{equation}

\subsection{The General Solution Theorem}

This short section contains the statement of our main theorem.

\begin{theorem}
\label{TheoGenSol}The general solution of eq. (\ref{TVAR(P)}) with free
constants (initial condition values) $y_{t-nl}$, $y_{t-nl-1},\ldots
,y_{t-nl-p+1}$ is given by 
\begin{equation}
y_{t,nl}^{gen}=y_{t,nl}^{hom}+y_{t,nl}^{par},  \label{TVAR(p)SOL}
\end{equation}%
where 
\begin{eqnarray*}
y_{t,nl}^{hom} &=&\xi _{t,nl}y_{t-nl}+\sum_{m=1}^{p-1}\dsum
\limits_{i=1}^{p-m}\phi _{m+i}(t-nl+i)\xi _{t,nl-i}y_{t-nl-m} \\
&=&\sum_{m=0}^{p-1}\dsum \limits_{i=1}^{p-m}\phi _{m+i}(t-nl+i)\xi
_{t,nl-i}y_{t-nl-m} \\
y_{t,nl}^{par} &=&\sum_{r=0}^{nl-1}\xi _{t,r}\phi
_{0}(t-r)+\sum_{r=0}^{nl-1}\xi _{t,r}\varepsilon _{t-r},
\end{eqnarray*}%
where the $\xi ${\footnotesize s} are expressed as generalized lower
continuant determinants (see eq. (\ref{KSIAR(p)})).
\end{theorem}

In the above Theorem $y_{t,nl}^{gen}$\ is decomposed into two parts: the $%
y_{t,nl}^{hom}$\ part, which consists of the $p$\ free constants ($%
y_{t-nl-m} $, $m=0,\ldots ,p-1$); and the $y_{t,nl}^{par}$\ part, which
contains the\ periodical drift terms ($\phi _{0}(\mathfrak{\cdot })$) and
the error terms ($\varepsilon ${\footnotesize s}) from time $t-nl+1$\ to
time $t$.

For `$n=0$' (for $i>j$ we use the convention $\sum_{r=i}^{j}(\cdot )=0$),
since $\xi _{t,0}=1$ and $\xi _{t,-c}=0$, $c>0$, (see eq. (\ref{KSIAR(p)})),
eq. (\ref{TVAR(p)SOL}) becomes an `identity': $y_{t,0}^{gen}=y_{t}$.
Similarly, when `$nl=1$' eq. (\ref{TVAR(p)SOL}), since $\xi _{t,1}=\phi
_{1}(t)$, $\xi _{t,0}=1$ and $\xi _{t,-c}=0$, $c>0$, it reduces to `eq. (\ref%
{TVAR(P)})': $y_{t,1}^{gen}=\sum_{m=1}^{p}\phi _{m}(t)y_{t+1-k-m}+\phi
_{0}(t)+\epsilon _{t}$.

Finally, for the PARMA($p,q;l$) model, which is given by%
\begin{equation}
\Phi _{t}(B)y_{t}=\phi _{0}(t)+\Theta _{t}(B)\varepsilon _{t},  \label{PARMA}
\end{equation}%
with $\Theta _{t}(B)=\dsum \nolimits_{j=1}^{q}\theta _{j}(t)B^{j}$, $\theta
_{j}(t)=\theta _{j}(\tau _{n})$, $j=1,\ldots ,q$, we replace $\varepsilon
_{t}$ in Theorem \ref{TheoGenSol} by $u_{t}=\Theta _{t}(B)\varepsilon _{t}$.

\section{OPTIMAL FORECASTING\label{SecSecMom}}

Having specified and solved a PARMA model by employing a univariate time
varying approach, we proceed to predict the future values of a periodically
correlated time series variable. Failure to allow for features of the data,
like seasonality, is likely to produce inferior forecasts. Accordingly, we
incorporate the non-stationarity of our series in a systematic manner into a
forecasting method. We begin by deriving the $nl$ step-ahead optimal linear
predictor.

Taking the conditional expectation of eq. (\ref{TVAR(p)SOL}) with respect to
the $\sigma $ field $\tciFourier _{\tau _{n}}$ ($\tau _{n}=t-nl$) yields the
following Proposition.

\begin{proposition}
\label{ProOptPred}For the PAR($p;l$) model the $nl$-step-ahead optimal (in $%
L_{2}$-sense) linear predictor of $y_{t}$, $\mathbb{E}(y_{t}\left \vert
\tciFourier _{\tau _{n}}\right. )$,\ is%
\begin{eqnarray}
\mathbb{E}(y_{t}\left \vert \tciFourier _{\tau _{n}}\right. )
&=&\sum_{r=0}^{nl-1}\xi _{t,r}\phi _{0}(t-r)+\xi
_{t,nl}y_{t-nl}+\sum_{m=1}^{p-1}\dsum \limits_{i=1}^{p-m}\phi
_{m+i}(t-nl+i)\xi _{t,nl-i}y_{t-nl-m}  \label{CE-TVAR(P)} \\
&=&\sum_{r=0}^{l-1}\sum_{T=0}^{n-1}\xi _{t,Tl+r}\phi _{0}(t-r)+\xi
_{t,nl}y_{t-nl}+\sum_{m=1}^{p-1}\dsum \limits_{i=1}^{p-m}\phi
_{m+i}(t-nl+i)\xi _{t,nl-i}y_{t-nl-m}.  \notag
\end{eqnarray}%
Next we consider the issue of forecast accuracy by examining the forecast
error resulting from the predictor. In particular, the forecast error for
the above $nl$-step-ahead predictor, $\mathbb{FE}(y_{t}\left \vert
\tciFourier _{\tau _{n}}\right. )=y_{t}-\mathbb{E}[y_{t}\left \vert
\tciFourier _{\tau _{n}}\right. ]$, is given by%
\begin{equation}
\mathbb{FE}(y_{t}\left \vert \tciFourier _{\tau _{n}}\right. )=\Xi
_{t,nl}(B)\varepsilon _{t}=\sum_{r=0}^{nl-1}\xi _{t,r}B^{r}\varepsilon _{t}.
\label{FE-TVAR(P)}
\end{equation}%
The optimal forecast is the one with the minimum square error; we provide
the following interpretation of the criterion. \ The multistep ahead
prediction error is expressed in terms of $nl$ error terms from time $t-nl+1$%
\ to time $t$ where the coefficient of the error term at time $t-r$, $\xi
_{t,r}$, is the determinant of an $r\times r$ matrix ($\Phi _{t,r}$), each
nonzero variable diagonal of which consists of the AR periodical
coefficients $\phi _{m}(\mathfrak{\cdot })$, $m=1,\ldots ,$min$(p,r)$\ from
time $t-r+m$\ to $t$. \newline
The mean square error is given by%
\begin{equation}
\mathbb{V}ar[\mathbb{FE(}y_{t}\left \vert \tciFourier _{\tau _{n}}\right.
)]=\Xi _{t,nl}^{(2)}(B)\sigma _{t}^{2}=\sum_{r=0}^{nl-1}\xi _{t,r}^{2}B^{r}%
\mathbb{\sigma }_{t}^{2}.  \label{VFE-TVAR(P)}
\end{equation}%
This error is expressed in terms of $nl$\ variances from time $t-nl+1$\ to
time $t$, with time varying coefficients (the squared $\xi ${\footnotesize s}
).%
\end{proposition}

\begin{remark}
\label{RemPARMA}For the PARMA($p,q;l$) model: $\mathbb{E}(y_{t}^{(\text{ARMA}%
)}\left \vert \tciFourier _{\tau _{n}}\right. )=\mathbb{E}(y_{t}\left \vert
\tciFourier _{\tau _{n}}\right. )+\sum_{r=nl}^{nl-1+q}\xi _{t,r}^{\prime
}\varepsilon _{t-r}$, where $\xi _{t,r}^{\prime }=\tsum
\nolimits_{j=r-nl+1}^{q}\xi _{t,r-j}\theta _{j}(t-r+j)$. $\mathbb{FE}%
(y_{t}^{(\text{ARMA})}\left \vert \tciFourier _{\tau _{n}}\right.
)=\sum_{r=0}^{nl-1}\xi _{t,r}^{\ast }\varepsilon _{t-r}$ where $\xi
_{t,r}^{\ast }=\xi _{t,r}+\tsum \nolimits_{j=1}^{q}\xi _{t,r-j}\theta
_{j}(t-r+j)$, and $\mathbb{V}ar[\mathbb{FE(}y_{t}^{(\text{ARMA})}\left \vert
\tciFourier _{\tau _{n}}\right. )]=\sum_{r=0}^{nl-1}(\xi _{t,r}^{\ast })^{2}%
\mathbb{\sigma }_{t-r}^{2}$.
\end{remark}

Using the vector season representation (see Appendix B) forecasts and
forecast error variances for a PARMA($p,q;l$) process can be computed. In
this manner we can construct forecasts as in $l$-variate VARMA($P,Q$) models
(see Ula, 1993). For example, Franses (1996a) and Franses and Paap (2005)
derive multi-step forecast error variances for low-order PAR models with $%
l=4 $, using the VS representation. \ But if $l$ is large, even low order
specifications will have large VAR representations and this is a handicap,
especially for forecasting. In contrast, our formulas using the univariate
framework allow a fast computation of the multi-step-ahead predictors even
if $l$ is large.

In what follows we give conditions for the first and second unconditional
moments of the model in eq. (\ref{TVAR(P)}) to exist.\qquad

\textbf{Assumption} 1. $\sum_{r=0}^{nl}\xi _{t,r}\phi _{0}(t-r)$ as $%
n\rightarrow \infty $ converges $\forall $ $t$ and $\tsum
\nolimits_{r=0}^{\infty }\sup_{t}(\xi _{t,r}^{2}\sigma _{t-r}^{2})<M<\infty $%
, $M\in \mathbb{R}^{+}$.

Assumption 1\ is a sufficient condition for the model in eq. (\ref{TVAR(P)})
to admit a second-order MA($\infty $) representation. A necessary but not
sufficient condition for $\sum_{r=0}^{nl}\xi _{t,r}\phi _{0}(t-r)$ to
converge is $\lim_{n\rightarrow \infty }[\xi _{t,nl}\phi _{0}(t-nl)]=0$ for
all $t$. A sufficient condition for this limit to be zero is: $%
\lim_{n\rightarrow \infty }\xi _{t,nl}=0$ and $\phi _{0}(t-nl)$ is bounded
with respect to $n$ $\forall $ $t$.

Pagano (1978) and Troutman (1979) were the first to study moment estimates
for PAR models and established their consistency and asymptotic efficiency.
\ Consistency and asymptotic efficiency of PARMA processes in the context of
least squares and maximum likelihood were established by Basawa and Lund
(2001).

Another consequence of Theorem \ref{TheoGenSol}\ are the following
Propositions, where we state expressions for the first two unconditional
moments of $y_{t}$. \ In the sequel we study the equivalent of the Wold
decomposition for non-stationary periodic processes. The challenge we face
is that in the periodical models we can not invert the AR polynomial due to
the presence of time dependent coefficients. We overcome this difficulty and
formulate a type of time varying Wold decomposition theorem.

\begin{proposition}
\label{ProWoldRepr}Let Assumption 1 hold. Then 
\begin{equation}
y_{t}\overset{L_{2}}{=}\lim_{n\rightarrow \infty }y_{t,nl}^{par}\overset{%
L_{2}}{=}\Xi _{t,\infty }(B)[\phi _{0}(t)+\varepsilon
_{t}]=\sum_{r=0}^{\infty }\xi _{t,r}B^{r}[\phi _{0}(t)+\varepsilon _{t}],
\label{TVAR(P)IMA}
\end{equation}%
is a unique solution of the PAR model in eq. (\ref{TVAR(P)}). The above
expression states that $\{y_{t,nl}^{par},t\in \mathbb{Z}\}$ (defined in eq. (%
\ref{TVAR(p)SOL})) $L_{2}$ converges as $n\rightarrow \infty $ if and only
if $\sum_{r=0}^{nl}\xi _{t,r}\phi _{0}(t-r)$ converges and $%
\sum_{r=0}^{nl}\xi _{t,r}\varepsilon _{t-r}$ converges a.s., and thus under
Assumption 1 $y_{t}\overset{L_{2}}{=}\lim_{n\rightarrow \infty
}y_{t,nl}^{par}$ satisfies eq. (\ref{TVAR(P)}). \newline
In other words $y_{t}$ is decomposed into a non random part%
%
\begin{equation}
\mathbb{E}(y_{t})=\lim_{n\rightarrow \infty }\mathbb{E}(y_{t}\left \vert
\tciFourier _{\tau _{n}}\right. )=\Xi _{t,\infty }(B)\phi
_{0}(t)=\sum_{r=0}^{\infty }\xi _{t,r}B^{r}\phi _{0}(t),  \label{E-TVAR(P)}
\end{equation}%
that is, an infinite sum of the periodical drifts where the time varying
coefficients are expressed as determinants of generalized lower continuant
matrices (the $\xi ${\footnotesize s}); and a zero mean random part%
\begin{equation*}
\lim_{n\rightarrow \infty }\mathbb{FE}(y_{t}\left \vert \tciFourier _{\tau
_{n}}\right. )=\sum_{r=0}^{\infty }\xi _{t,r}B^{r}\varepsilon _{t}.
\end{equation*}%
Therefore, the $\xi _{t,r}$\ as defined in eq. (\ref{KSIAR(p)})
are the Green functions associated with $\Phi _{t}(B)$\ (see also
Paraskevopoulos, Karanasos, and Dafnos, 2013). For the PARMA($p,q;l$) model
we replace $\varepsilon _{t-r}$ by $u_{t-r}=\Theta _{t}(B)\varepsilon _{t-r}$
or $\xi _{t,r}$ by $\xi _{t,r}^{\ast }$ (see Remark \ref{RemPARMA}).
\end{proposition}

Tests to detect periodicities in the autocovariances of a realized series
have been proposed by, among others, Vecchia and Ballerini (1991).

Next we state as a Proposition the result for the second moment structure.

\begin{proposition}
\label{ProSecMom}Let Assumption 1 hold. Then the second unconditional moment
for the PAR($p;l$) model exists and it is given by%
\begin{equation}
\mathbb{E}(y_{t}^{2})=[\mathbb{E}(y_{t})]^{2}+\Xi _{t,\infty
}^{(2)}(B)\sigma _{t}^{2}=[\mathbb{E}(y_{t})]^{2}+\sum_{r=0}^{\infty }\xi
_{t,r}^{2}B^{r}\sigma _{t}^{2}.  \label{Var-TVAR(P)}
\end{equation}%
That is, the time varying variance of $y_{t}$ is an infinite sum of the time
varying variances of the errors with time varying coefficients (the squared
values of the $\xi ${\footnotesize s}). \newline
In addition, the time varying autocovariance function $\gamma _{t,nl}$ is
given by 
\begin{eqnarray}
\gamma _{t,nl} &=&\mathbb{C}ov(y_{t},y_{\tau _{n}})=\sum_{r=0}^{\infty }\xi
_{t,nl+r}\xi _{\tau _{n},r}\sigma _{\tau _{n}-r}^{2}=\xi _{t,nl}\mathbb{V}%
ar(y_{\tau _{n}})+  \label{Covar-TVAR(P)} \\
&&\sum_{m=1}^{p-1}\dsum \limits_{i=1}^{p-m}\phi _{m+i}(\tau _{n}+i)\xi
_{t,nl-i}\mathbb{C}ov(y_{\tau _{n}},y_{\tau _{n}-m}),  \notag
\end{eqnarray}%
where the second equality follows from the MA($\infty $) representation of $%
y_{t}$ in eq. (\ref{TVAR(P)IMA}) and the third one from eq. (\ref{TVAR(p)SOL}%
) in Theorem \ref{TheoGenSol}. For any fixed $t$, $\lim_{n\rightarrow \infty
}\gamma _{t,nl}=0$ when $\lim_{n\rightarrow \infty }\xi _{t,nl}=0$. \newline
Finally, for the PARMA($p,q;l$) model we replace the $\xi ${\footnotesize s}
in eq. (\ref{Var-TVAR(P)}) and in the second equality in eq. (\ref%
{Covar-TVAR(P)}) by the $\xi ^{\ast }${\footnotesize s} (as defined in
Remark \ \ref{RemPARMA}), and we add the term $\sum_{r=0}^{q-1}\xi
_{t-nl,r}^{\ast }\xi _{t,r+nl}^{\prime }\sigma _{t-nl-r}^{2}$ in the third
equality of eq.(\ref{Covar-TVAR(P)}).
\end{proposition}

Although it may be difficult to explicitly compute the covariance structure
of $\{y_{t}\}$, for numerical work, one can always calculate it by computing
the Green functions (that is, the continuant determinants $\xi $%
{\footnotesize s}) with eqs. (\ref{FAIMAR(p)X}) and (\ref{KSIAR(p)}) and
adding them up with eq. (\ref{Covar-TVAR(P)}).

\section{CONCLUSIONS}

We have presented a univariate TV treatment of the periodic ARMA model. \ We
have provided the general solution for the $p$th order periodic linear
stochastic process, as the sum of the homogeneous and particular solutions,
both expressed in terms of Hessenbergians. The solution is derived from a
general method for solving infinite linear systems in row-finite form, which
employs the infinite Gaussian part of the Gauss-Jordan algorithm.

Several advantages are associated with our approach. \ We are able to
examine a single seasonal time series with a univariate framework. \ The
large number of variables that might be involved in the vector
representation line of research that follows Gladyshev (1961) will be a
handicap, particularly for forecasting. The parsimonious character of our
modeling is especially useful when it comes to applying periodic processes
to daily and high frequency data. In addition, freeing ourselves from the
bounds of recursion, which lies behind the vector treatment of
periodicities, we have been able to provide explicit formulas for optimal
predictors and for the second moment structure.

Our results include those for the ARMA model with constant coefficients as a
special case. They are also extendible to the solutions of infinite and
ascending order specifications. One natural extension of our paper is to
apply the univariate methodology to multivariate seasonal models, that is to
treat not only a single seasonal time series but multiple series as well,
each, with a univariate framework.

\bigskip

REFERENCES

\bigskip

Anderson, P.L., M.M. Meerschaert, \& K. Zhang (2013) Forecasting with
prediction intervals for periodic autoregressive moving average models. 
\textit{Journal of Time Series Analysis} 34, 187--193.

Basawa, I.V. \& R. Lund (2001) Large sample properties of parameter
estimates for periodic ARMA models. \textit{Journal of Time Series Analysis}
22, 651--663.

Bollerslev, T. \& E. Ghysels (1996) Periodic autoregressive conditional
heteroscedasticity. \textit{Journal of Business \& Economic Statistics} 14,
139--151.

Cleveland, W.P. \& G.C. Tiao (1979) Modeling seasonal time series. \textit{%
Revue Economic Appliqu\'{e}e} 32, 107--129.

del Barrio Castro, T. \& D.R. Osborn (2008) Testing for seasonal unit roots
in periodic integrated autoregressive processes. \textit{Econometric Theory}
24, 1093--1129.

Franses, P.H. (1991) Seasonality, non-stationarity, and the forecasting of
monthly time series. \textit{International Journal of Forecasting} 7,
199--208.

Franses, P.H. (1994) A multivariate approach to modeling univariate seasonal
time series. \textit{Journal of Econometrics 63}, 133--51.

Franses, P.H. (1996a) Multi-step forecast error variances for periodically
integrated time series. \textit{Journal of Forecasting} 15, 83--95.

Franses, P.H. (1996b) \textit{Periodicity and Stochastic Trends in Economic
Time Series}, Oxford University Press.

Franses, P.H. \& R. Paap (2004) \textit{Periodic Time Series Models}, Oxford
University Press.

Franses, P.H. \& R. Paap (2005) Forecasting with periodic autoregressive
time-series models. In M.P. Clements \& D.F. Hendry (eds.), \textit{A
Companion to Economic Forecasting}, pp. 432--452. Wiley-Blackwell.

Ghysels, E. (1988) A Study toward a dynamic theory of seasonality for
economic time series. \textit{Journal of the American Statistical Association%
} 83, 168--172.

Ghysels, E. \& D.R. Osborn (2001) \textit{The Econometric Analysis of
Seasonal Time Series}, Cambridge University Press.

Gladyshev, E.G. (1961) On periodically correlated random sequences. \textit{%
Soviet Mathematics} 2, 385--388.

Jones, R.H. \& W.M. Brelsford (1967) Time series with periodic structure. 
\textit{Biometrika} 54, 403--408.

Hurd, H.L. \& A. Miamee (2007) \textit{Periodically Correlated Random
Sequences: Spectral Theory and Practice}, Wiley-Blackwell.

Karanasos, M., A.G. Paraskevopoulos, F. Menla Ali, M. Karoglou, \& M. Yfanti
(2014) Modeling returns and volatilities during financial crises: a time
varying coefficient approach. \textit{Journal of Empirical Finance},
forthcoming.

Lund, R. \& I.V. Basawa (2000) Recursive prediction and likelihood
evaluation for periodic ARMA models. \textit{Journal of Time Series Analysis}
21, 75--93.

Lund, R., Q. Shao, \& I. Basawa (2006) Parsimonious periodic time series
modeling. \textit{Australian \& New Zealand Journal of Statistics} 48,
33--47.

Miron, J.A. (1986) Seasonal fluctuations and the life cycle-permanent income
model of consumption. \textit{Journal of Political Economy} 94, 1258--1279.

Osborn, D.R. (1988) Seasonality and habit persistence in a life cycle model
of consumption. \textit{Journal of Applied Econometrics} 3, 255--266.

Osborn, D.R. (1990) A survey of seasonality in UK macroeconomic variables. 
\textit{International Journal of Forecasting} 6, 327--336.

Osborn, D.R. (1991) The implications of periodically varying coefficients
for seasonal time-series processes. \textit{Journal of Econometrics} 48,
373--384.

Pagano, M. (1978) On periodic and multiple autoregressions. \textit{The} 
\textit{Annals of Statistics} 6, 1310--1317.

Paraskevopoulos, A.G. (2012) The Infinite Gauss-Jordan elimination on
row-finite $\omega \times \omega $ matrices. arXiv: 1201.2950.

Paraskevopoulos, A.G., M. Karanasos, \& S. Dafnos (2013) A unified theory
for time varying models: foundations with applications in the presence of
breaks and heteroskedasticity (and some results on companion and Hessenberg
matrices). \textit{Unpublished Paper}.

Parzen, E. \& M. Pagano (1979) An approach to modeling seasonally stationary
time series. \textit{Journal of Econometrics} 9, 137--153.

Shao, Q. (2008) Robust estimation for periodic autoregressive time series. 
\textit{Journal of Time Series Analysis} 29, 251--263.

Taylor, A.M.R. (2002) Regression-based unit root tests with recursive mean
adjustment for seasonal and nonseasonal time series. \textit{Journal of
Business \& Economic Statistics} 20, 269--281.

Tesfaye, Y.G., P.L. Anderson, \& M.M. Meerschaert (2011) Asymptotic results
for Fourier-PARMA time series. \textit{Journal of Time Series Analysis} 32,
157-174.

Tiao, G.C. \& M.R. Grupe (1980) Hidden periodic autoregressive-moving
average models in time series data. \textit{Biometrika} 67, 365--373.

Tiao, G.C. \& I. Guttman (1980) Forecasting contemporal aggregates of
multiple time series. \textit{Journal of Econometrics} 12, 219--230.

Troutman, B.M. (1979) Some results in periodic autoregression. \textit{%
Biometrika} 66, 219-228.

Ula, T.A. (1993) Forecasting of multivariate periodic autoregressive
moving-average processes. \textit{Journal of Time Series Analysis} 14,
645--657.

Vecchia, A.V. (1985) Periodic autoregressive-moving average (PARMA) modeling
with applications to water resources. \textit{Journal of the American Water
Resources Association }21, 721-730.

Vecchia, A.V. \& R. Ballerini (1991) Testing for periodic autocorrelations
in seasonal time series data. \textit{Biometrika} 78, 53--63.

\appendix

\section{APPENDIX}

\begin{proof}
(Theorem 1; for the PARMA($p,q;l$) model in eq. (\ref{PARMA})). We will
denote the $nl\times (p+nl)$ top submatrix of $\mathbf{\Phi }$ (associated
with eq. (\ref{PARMA})) by $\mathbf{A}_{t,\tau _{n}}$(recall that $\tau
_{n}=t-nl$)%
\begin{eqnarray*}
&&\mathbf{A}_{t,\tau _{n}}%
\begin{tabular}{l}
=%
\end{tabular}
\\
&&\left( 
\begin{array}{ccccccccccc}
\phi _{p}(\tau _{n}+1) & \phi _{p-1}(\tau _{n}+1) & \cdots & \phi _{1}(\tau
_{n}+1) & -1 &  &  &  &  &  &  \\ 
& \phi _{p}(\tau _{n}+2) & \cdots & \phi _{2}(\tau _{n}+2) & \phi _{1}(\tau
_{n}+2) & \ddots &  &  &  &  &  \\ 
&  & \cdots & \vdots & \vdots & \ddots & \ddots &  &  &  &  \\ 
&  &  &  & \phi _{p}(\tau _{n}+p+1) & \ddots & \ddots & \ddots &  &  &  \\ 
&  &  &  &  & \ddots & \cdots & \ddots & \ddots &  &  \\ 
&  &  &  &  &  & \phi _{p}(t-1) & \cdots & \phi _{1}(t-1) & -1 &  \\ 
&  &  &  &  &  &  & \cdots & \phi _{2}(t) & \phi _{1}(t) & -1%
\end{array}%
\right) .
\end{eqnarray*}%
In view of eq. (\ref{PARMA}) we define the forcing term $r_{t}=\phi
_{0}(t)+u_{t}$ (recall that $u_{t}=\varepsilon _{t}+\sum_{j=1}^{q}\theta
_{j}(t)B^{j}\varepsilon _{t}$) along with the $nl\times 1$ vector $\mathbf{r}%
_{t,\tau _{n}}=(r_{\tau _{n}+1},r_{\tau _{n}+2},...,r_{t})^{\prime }$. Eq. (%
\ref{PARMA}) can be written as 
\begin{equation*}
y_{t}-\sum_{m=1}^{p}\phi _{m}(t)y_{t-m}=r_{t}.
\end{equation*}%
The solution $(p+nl)\times 1$ vector $\mathbf{y}_{t,\tau _{n}}=(y_{\tau
_{n}-p+1},y_{\tau _{n}-p+2},...,y_{\tau _{n}},y_{\tau _{n}+1,1},...,y_{\tau
_{n}+nl,nl})^{\prime }$ of the overdetermined system 
\begin{equation}
\mathbf{A}_{t,\tau _{n}}\cdot \mathbf{y}_{t,\tau _{n}}=\mathbf{r}_{t,\tau
_{n}}  \tag{A.1}
\end{equation}%
contains the $p$ free constants followed by the first $nl=t-\tau _{n}$
solutions of eq. (\ref{PARMA}). Let us call $\mathbf{e}_{k}=(0,0,...,0,1)^{%
\prime }$\vspace{0.1in} the $k$th unit vector of the canonical basis of $%
\mathbb{R}^{k}$. We introduce the matrix $\tilde{\mathbf{C}}_{t,\tau _{n}}$
consisting of the core solution matrix $\mathbf{C}_{t,nl}$ (see eq. (\ref%
{CoreMatrix})) augmented by the column vector $-\mathbf{e}_{nl}$: 
\begin{equation*}
\tilde{\mathbf{C}}_{t,\tau _{n}}=\left( 
\begin{array}{ccccccccc}
-1 &  &  &  &  &  &  &  &  \\ 
\phi _{1}(\tau _{n}+2) & -1 &  &  &  &  &  &  &  \\ 
\phi _{2}(\tau _{n}+3) & \phi _{1}(\tau _{n}+3) & \ddots &  &  &  &  &  & 
\\ 
\vdots & \vdots & \ddots & \ddots &  &  &  &  &  \\ 
\phi _{p}(\tau _{n}+p+1) & \phi _{p-1}(\tau _{n}+p+1) & \ddots & \ddots & 
\ddots &  &  &  &  \\ 
& \phi _{p}(\tau _{n}+p+2) & \ddots & \ddots & \ddots & \ddots &  &  &  \\ 
&  & \ddots & \ddots & \ddots & \ddots & \ddots &  &  \\ 
&  &  & \phi _{p}(t-1) & \phi _{p-1}(t-1) & \cdots & \phi _{1}(t-1) & -1 & 
\\ 
&  &  &  & \phi _{p}(t) & \cdots & \phi _{2}(t) & \phi _{1}(t) & -1%
\end{array}%
\right) .
\end{equation*}%
Evidently $\tilde{\mathbf{C}}_{t,\tau _{n}}$ is a $nl\times nl$ nonsingular
submatrix of $\mathbf{A}_{t,\tau _{n}}$. The matrix $\mathbf{A}_{t,\tau
_{n}} $ is partitioned into two submatrices:

\begin{itemize}
\item The matrix $\tilde{\mathbf{C}}_{t,\tau _n}$ and the

\item $nl\times p$ matrix 
\begin{equation*}
\mathbf{P}_{t,\tau _{n}}=\left( 
\begin{array}{cccc}
\phi _{p}(\tau _{n}+1) & \phi _{p-1}(\tau _{n}+1) & \cdots & \phi _{1}(\tau
_{n}+1) \\ 
0 & \phi _{p}(\tau _{n}+1) & \cdots & \phi _{2}(\tau _{n}+2) \\ 
0 & 0 & \cdots & \phi _{3}(\tau _{n}+3) \\ 
\vdots & \vdots & \vdots \vdots \vdots & \vdots \\ 
0 & 0 & \cdots & \phi _{p}(\tau _{n}+p) \\ 
0 & 0 & \cdots & 0 \\ 
\vdots & \vdots & \vdots \vdots \vdots & \vdots \\ 
0 & 0 & \cdots & 0%
\end{array}%
\right) .
\end{equation*}
\end{itemize}

Therefore the system (A.1) can be equivalently expressed as 
\begin{equation*}
(\mathbf{P}_{t,\tau _{n}}|\tilde{\mathbf{C}}_{t,\tau _{n}})\cdot \mathbf{y}%
_{t,\tau _{n}}=\mathbf{r}_{t,\tau _{n}}.
\end{equation*}%
Block matrix multiplication entails that 
\begin{equation*}
\begin{array}{l}
(\mathbf{P}_{t,\tau _{n}}|\tilde{\mathbf{C}}_{t,\tau _{n}})\left( 
\begin{array}{c}
y_{\tau _{n}-p+1} \\ 
y_{\tau _{n}-p+2}\vspace{-0.05in} \\ 
\vdots \\ 
y_{\tau _{n}}\vspace{0.1in} \\ \hline
y_{\tau _{n}+1,1}\vspace{-0.1in} \\ 
y_{\tau _{n}+2,2} \\ 
\vdots \\ 
y_{t,nl}%
\end{array}%
\right) =\left( 
\begin{array}{c}
r_{\tau _{n}+1} \\ 
r_{\tau _{n}+2} \\ 
\vdots \\ 
r_{t}%
\end{array}%
\right) \Longleftrightarrow \vspace{0.15in} \\ 
\mathbf{P}_{t,\tau _{n}}\left( \! \! \!%
\begin{array}{c}
y_{\tau _{n}-p+1} \\ 
y_{\tau _{n}-p+2}\vspace{-0.05in} \\ 
\vdots \\ 
y_{\tau _{n}}\vspace{0.1in}%
\end{array}%
\! \! \! \right) +\tilde{\mathbf{C}}_{t,\tau _{n}}\cdot \left( \! \! \!%
\begin{array}{c}
y_{\tau _{n}+1,1} \\ 
y_{\tau _{n}+2,2}\vspace{-0.05in} \\ 
\vdots \\ 
y_{t,nl}%
\end{array}%
\! \! \! \right) =\left( 
\begin{array}{c}
r_{\tau _{n}+1} \\ 
r_{\tau _{n}+2} \\ 
\vdots \\ 
r_{t}%
\end{array}%
\right) ,%
\end{array}%
\end{equation*}%
whence 
\begin{equation}
\tilde{\mathbf{C}}_{t,\tau _{n}}\cdot \left( \! \! \!%
\begin{array}{c}
y_{\tau _{n}+1,1} \\ 
y_{\tau _{n}+2,2}\vspace{-0.05in} \\ 
\vdots \\ 
y_{t,nl}%
\end{array}%
\! \! \! \right) =-\mathbf{P}_{t,\tau _{n}}\cdot \left( \! \! \!%
\begin{array}{c}
y_{\tau _{n}-p+1} \\ 
y_{\tau _{n}-p+2}\vspace{-0.05in} \\ 
\vdots \\ 
y_{\tau _{n}}\vspace{0.1in}%
\end{array}%
\! \! \! \right) +\left( 
\begin{array}{c}
r_{\tau _{n}+1} \\ 
r_{\tau _{n}+2} \\ 
\vdots \\ 
r_{t}%
\end{array}%
\right) .  \tag{A.2}
\end{equation}%
Employing the notation 
\begin{equation*}
h_{m}=\sum_{j=m}^{p}\phi _{p-j+m}(\tau _{n}+m)y_{\tau _{n}-p+j},\
m=1,2,\ldots ,p,
\end{equation*}%
the right hand side of (A.2) takes the form: 
\begin{equation*}
\begin{array}{l}
-\left( 
\begin{array}{cccc}
\phi _{p}(\tau _{n}+1) & \phi _{p-1}(\tau _{n}+1) & \cdots & \phi _{1}(\tau
_{n}+1) \\ 
0 & \phi _{p}(\tau _{n}+2) & \cdots & \phi _{2}(\tau _{n}+2) \\ 
0 & 0 & \cdots & \phi _{3}(\tau _{n}+3) \\ 
\vdots & \vdots & \vdots \vdots \vdots & \vdots \\ 
0 & 0 & \cdots & \phi _{p}(\tau _{n}+p) \\ 
0 & 0 & \cdots & 0 \\ 
\vdots & \vdots & \vdots \vdots \vdots & \vdots \\ 
0 & 0 & \cdots & 0%
\end{array}%
\right) \cdot \left( \! \! \!%
\begin{array}{c}
y_{\tau _{n}-p+1} \\ 
y_{\tau _{n}-p+2}\vspace{-0.05in} \\ 
\vdots \\ 
y_{\tau _{n}}\vspace{0.1in}%
\end{array}%
\! \! \! \right) +\left( 
\begin{array}{c}
r_{\tau _{n}+1} \\ 
r_{\tau _{n}+2} \\ 
\vdots \\ 
r_{t}%
\end{array}%
\right) = \\ 
\left( 
\begin{array}{c}
\displaystyle-\sum_{j=1}^{p}\phi _{p-j+1}(\tau _{n}+1)y_{\tau _{n}-p+j}%
\vspace{0.1in} \\ 
\displaystyle-\sum_{j=2}^{p}\phi _{p-j+2}(\tau _{n}+2)y_{\tau _{n}-p+j} \\ 
\vdots \\ 
-\phi _{p}(\tau _{n}+p)y_{\tau _{n}} \\ 
0 \\ 
\vdots \\ 
0%
\end{array}%
\right) +\left( 
\begin{array}{c}
r_{\tau _{n}+1} \\ 
r_{\tau _{n}+2} \\ 
\vdots \\ 
r_{\tau _{n}+p} \\ 
r_{\tau _{n}+p+1}\vspace{0.1in} \\ 
\vdots \\ 
r_{t}%
\end{array}%
\right) =\left( 
\begin{array}{c}
r_{\tau _{n}+1}-h_{1} \\ 
r_{\tau _{n}+2}-h_{2} \\ 
\vdots \\ 
r_{\tau _{n}+p}-h_{p} \\ 
r_{\tau _{n}+p+1}\vspace{0.1in} \\ 
\vdots \\ 
r_{t}%
\end{array}%
\right) .%
\end{array}%
\end{equation*}%
Thus (A.2) can be written as 
\begin{equation}
\tilde{\mathbf{C}}_{t,\tau _{n}}\cdot \left( \! \! \!%
\begin{array}{c}
y_{\tau _{n}+1,1} \\ 
y_{\tau _{n}+2,2} \\ 
\vdots \\ 
y_{\tau _{n}+p,p} \\ 
y_{\tau _{n}+p+1,p+1} \\ 
\vdots \\ 
y_{t,nl}%
\end{array}%
\! \! \! \right) =\left( 
\begin{array}{c}
r_{\tau _{n}+1}-h_{1} \\ 
r_{\tau _{n}+2}-h_{2} \\ 
\vdots \\ 
r_{\tau _{n}+p}-h_{p} \\ 
r_{\tau _{n}+p+1} \\ 
\vdots \\ 
r_{t}%
\end{array}%
\right) .  \tag{A.3}
\end{equation}%
As $\tilde{\mathbf{C}}_{t,\tau _{n}}$ is nonsingular the system (A.3) has a
unique solution. By Cramer's rule the general solution $%
y_{t,nl}^{gen}=y_{t,nl}$ is the fraction of two determinants: The numerator
is the determinant of the matrix $\tilde{\mathbf{C}}_{t,\tau _{n}}$ whose
last column is replaced by the right hand side column of (A.3), and the
denominator is $\det (\tilde{\mathbf{C}}_{t,\tau _{n}})$. Taking into
account that $\det (\tilde{\mathbf{C}}_{t,\tau _{n}})=(-1)^{nl}$ it follows
that 
\begin{equation*}
y_{t,nl}=\frac{\left \vert 
\begin{array}{cccccccc}
-1 &  &  &  &  &  &  & r_{\tau _{n}+1}-h_{1} \\ 
\phi _{1}(\tau _{n}+2) & \ddots &  &  &  &  &  & r_{\tau _{n}+2}-h_{2} \\ 
\vdots & \ddots & \ddots &  &  &  &  & \vdots \\ 
\phi _{p-1}(\tau _{n}+p) & \ddots & \ddots & \ddots &  &  &  & r_{\tau
_{n}+p}-h_{p} \\ 
\phi _{p}(\tau _{n}+p+1) & \ddots & \ddots & \ddots & \ddots &  &  & r_{\tau
_{n}+p+1} \\ 
& \ddots & \ddots & \ddots & \ddots & \ddots &  & \vdots \\ 
&  & \phi _{p}(t-1) & \phi _{p-1}(t-1) & \cdots & \phi _{1}(t-1) & -1 & 
r_{t-1} \\ 
&  &  & \phi _{p}(t) & \cdots & \phi _{2}(t) & \phi _{1}(t) & r_{t}%
\end{array}%
\right \vert }{(-1)^{nl}}.
\end{equation*}%
As a column exchange between two consecutive columns of a determinant
changes the sign of the determinant, we conclude that after $nl-1$ column
exchanges the last column moves to the first, yielding 
\begin{equation*}
y_{t,nl}=(-1)^{nl-1}\frac{\left \vert 
\begin{array}{cccccccc}
r_{\tau _{n}+1}-h_{1} & -1 &  &  &  &  &  &  \\ 
r_{\tau _{n}+2}-h_{2} & \phi _{1}(\tau _{n}+2) & \ddots &  &  &  &  &  \\ 
\vdots & \vdots & \ddots & \ddots &  &  &  &  \\ 
r_{\tau _{n}+p}-h_{p} & \phi _{p-1}(\tau _{n}+p) & \ddots & \ddots & \ddots
&  &  &  \\ 
r_{\tau _{n}+p+1} & \phi _{p}(\tau _{n}+p+1) & \ddots & \ddots & \ddots & 
\ddots &  &  \\ 
\vdots &  & \ddots & \ddots & \ddots & \ddots & \ddots &  \\ 
r_{t-1} &  &  & \phi _{p}(t-1) & \phi _{p-1}(t-1) & \cdots & \phi _{1}(t-1)
& -1 \\ 
r_{t} &  &  &  & \phi _{p}(t) & \cdots & \phi _{2}(t) & \phi _{1}(t)%
\end{array}%
\right \vert }{(-1)^{nl}},
\end{equation*}%
thus 
\begin{equation*}
y_{t,nl}=\left \vert \! \! \!%
\begin{array}{cccccccc}
h_{1}-r_{\tau _{n}+1} & -1 &  &  &  &  &  &  \\ 
h_{2}-r_{\tau _{n}+2} & \phi _{1}(\tau _{n}+2) & \ddots &  &  &  &  &  \\ 
\vdots & \vdots & \ddots & \ddots &  &  &  &  \\ 
h_{p}-r_{\tau _{n}+p} & \phi _{p-1}(\tau _{n}+p) & \ddots & \ddots & \ddots
&  &  &  \\ 
-r_{\tau _{n}+p+1} & \phi _{p}(\tau _{n}+p+1) & \ddots & \ddots & \ddots & 
\ddots &  &  \\ 
\vdots &  & \ddots & \ddots & \ddots & \ddots & \ddots &  \\ 
-r_{t-1} &  &  & \phi _{p}(t-1) & \phi _{p-1}(t-1) & \cdots & \phi _{1}(t-1)
& -1 \\ 
-r_{t} &  &  &  & \phi _{p}(t) & \cdots & \phi _{2}(t) & \phi _{1}(t)%
\end{array}%
\right \vert .
\end{equation*}

Using the definition of the forcing term $r_{t}$ we can write $y_{t,nl}$ as 
\begin{equation*}
y_{t,nl}=\left \vert \! \! \!%
\begin{array}{ccccccc}
h_{1}+\phi _{0}(\tau _{n}+1)+{\small u}_{\tau _{n}+1} & -1 &  &  &  &  &  \\ 
h_{2}+\phi _{0}(\tau _{n}+2)+{\small u}_{\tau _{n}+2} & \phi _{1}(\tau
_{n}+2) & \ddots &  &  &  &  \\ 
\vdots & \vdots & \ddots & \ddots &  &  &  \\ 
\phi _{0}(\tau _{n}+p+1)+{\small u}_{\tau _{n}+p+1} & \phi _{p}(\tau
_{n}+p+1) & \ddots & \ddots & \ddots &  &  \\ 
\vdots &  & \ddots & \ddots & \ddots & \ddots &  \\ 
\phi _{0}(t-1)+{\small u}_{t-1} &  &  & \phi _{p}(t-1) & \cdots & \phi
_{1}(t-1) & -1 \\ 
\phi _{0}(t)+{\small u}_{t} &  &  &  & \cdots & \phi _{2}(t) & \phi _{1}(t)%
\end{array}%
\right \vert .
\end{equation*}%
In the above formula we expressed the general solution ($%
y_{t,nl}^{gen}=y_{t,nl}$) as a Hessenbergian. Next we will decompose it into
two parts: the homogeneous and the particular solutions.\textbf{\ \ }%
Expanding the determinant along the first column we have: 
\begin{equation*}
\begin{array}{ll}
y_{t,nl}\! \! \! & =\displaystyle \sum_{m=1}^{p}(\phi _{0}(\tau _{n}+m)+%
{\small u}_{\tau _{n}+m}+h_{m})\xi _{t,nl-m}+\sum_{m=p+1}^{nl}(\phi
_{0}(\tau _{n}+m)+{\small u}_{\tau _{n}+m})\xi _{t,nl-m} \\ 
& =\displaystyle \sum_{m=1}^{p}(\phi _{0}(\tau _{n}+m)+{\small u}_{\tau
_{n}+m})\xi _{t,nl-m}+\sum_{m=1}^{p}h_{m}\xi
_{t,nl-m}+\sum_{m=p+1}^{nl}(\phi _{0}(\tau _{n}+m)+{\small u}_{\tau
_{n}+m})\xi _{t,nl-m} \\ 
& =\displaystyle \sum_{m=1}^{nl}(\phi _{0}(\tau _{n}+m)+{\small u}_{\tau
_{n}+m})\xi _{t,nl-m}+\sum_{m=1}^{p}h_{m}\xi _{t,nl-m} \\ 
& =\displaystyle \sum_{m=1}^{nl}(\phi _{0}(\tau _{n}+m)+{\small u}_{\tau
_{n}+m})\xi _{t,nl-m}+\sum_{m=1}^{p}\sum_{j=m}^{p}\phi _{p-j+m}(\tau
_{n}+m)y_{\tau _{n}-p+j}\xi _{t,nl-m}.%
\end{array}%
\end{equation*}%
The first sum in $y_{t,nl}$, is the particular solution $y_{t,nl}^{par}$,
and it can be written as 
\begin{equation*}
\begin{array}{ll}
\displaystyle \sum_{m=1}^{nl}(\phi _{0}(\tau _{n}+m)+{\small u}_{\tau
_{n}+m})\xi _{t,nl-m} & =\displaystyle \sum_{m=1}^{nl}(\phi _{0}(t-nl+m)+%
{\small u}_{t-nl+m})\xi _{t,nl-m} \\ 
& =\displaystyle \sum_{m=1}^{nl}(\phi _{0}(t-(nl-m))+{\small u}%
_{t-(nl-m)})\xi _{t,nl-m} \\ 
& =\displaystyle \sum_{r=0}^{nl-1}\xi _{t,r}\phi
_{0}(t-r)+\sum_{r=0}^{nl-1}\xi _{t,r}{\small u}_{t-r}.%
\end{array}%
\end{equation*}%
Next expand the second sum, 
\begin{equation*}
\begin{array}{l}
\displaystyle \sum_{m=1}^{p}\sum_{j=m}^{p}\phi _{p-j+m}(\tau _{n}+m)y_{\tau
_{n}-p+j}\xi _{t,nl-m}=\vspace{0.1in} \\ 
\lbrack \phi _{1}(\tau _{n}+1)y_{\tau _{n}}+\phi _{2}(\tau _{n}+1)y_{\tau
_{n}-1}+...+\phi _{p-1}(\tau _{n}+1)y_{\tau _{n}-p+2}+\phi _{p}(\tau
_{n}+1)y_{\tau _{n}-p+1}]\xi _{t,nl-1} \\ 
+[\phi _{2}(\tau _{n}+2)y_{\tau _{n}}+\phi _{3}(\tau _{n}+2)y_{\tau
_{n}-1}+...+\phi _{p}(\tau _{n}+2)y_{\tau _{n}-p+2}]\xi _{t,nl-2}+...+ \\ 
\lbrack \phi _{p-1}(\tau _{n}+p-1)y_{\tau _{n}}+\phi _{p}(\tau
_{n}+p-1)y_{\tau _{n}-1}]\xi _{t,nl-p+1}+\phi _{p}(\tau _{n}+p)y_{\tau
_{n}}\xi _{t,nl-p}.%
\end{array}%
\end{equation*}%
Factoring the above expansion relative to $y${\footnotesize s} we get the
final form of the homogeneous solution in terms of the initial conditions, $%
y_{t,nl}^{\hom }$, 
\begin{equation*}
\begin{array}{ll}
\left[ \phi _{1}(\tau _{n}+1)\xi _{t,nl-1}+\phi _{2}(\tau _{n}+2)\xi
_{t,nl-2}+\ldots +\phi _{p-1}(\tau _{n}+p-1)\xi _{t,nl-p+1}+\phi _{p}(\tau
_{n}+p)\xi _{t,nl-p}\right] y_{\tau _{n}}+ &  \\ 
\left[ \phi _{2}(\tau _{n}+1)\xi _{t,nl-1}+\phi _{3}(\tau _{n}+2)\xi
_{t,nl-2}+...+\phi _{p}(\tau _{n}+p-1)\xi _{t,nl-p+1}\right] y_{\tau
_{n}-1}+...+ &  \\ 
\left[ \phi _{p-1}(\tau _{n}+1)\xi _{t,nl-1}+\phi _{p}(\tau _{n}+2)\xi
_{t,nl-2}\right] y_{\tau _{n}-p+2}+\phi _{p}(\tau _{n}+1)\xi
_{t,nl-1}y_{\tau _{n}-p+1}=\vspace{0.15in} &  \\ 
\displaystyle \sum_{m=0}^{p-1}\left( \sum_{i=1}^{p-m}\phi _{m+i}(\tau
_{n}+i)\xi _{t,nl-i}\right) y_{\tau _{n}-m}. & 
\end{array}%
\end{equation*}%
Expanding the determinant of $\Phi _{t,nl}$, that is $\xi _{t,nl}$, along
the first column we have%
\begin{equation*}
\xi _{t,nl}=\phi _{1}(\tau _{n}+1)\xi _{t,nl-1}+\phi _{2}(\tau _{n}+2)\xi
_{t,nl-2}+\ldots +\phi _{p-1}(\tau _{n}+p-1)\xi _{t,nl-p+1}+\phi _{p}(\tau
_{n}+p)\xi _{t,nl-p},
\end{equation*}%
and thus we can also write%
\begin{equation*}
\sum_{m=0}^{p-1}\left( \sum_{i=1}^{p-m}\phi _{m+i}(\tau _{n}+i)\xi
_{t,nl-i}\right) y_{\tau _{n}-m}=\xi
_{t,nl}y_{t-nl}+\sum_{m=1}^{p-1}\sum_{i=1}^{p-m}\phi _{m+i}(t-nl+i)\xi
_{t,nl-i}y_{t-nl-m}.
\end{equation*}%
Accordingly the general solution is given by 
\begin{eqnarray*}
y_{t,nl} &=&\xi _{t,nl}y_{t-nl}+\sum_{m=1}^{p-1}\sum_{i=1}^{p-m}\phi
_{m+i}(t-nl+i)\xi _{t,nl-i}y_{t-nl-m}+\sum_{r=0}^{nl-1}\xi _{t,r}\phi
_{0}(t-r)+\sum_{r=0}^{nl-1}\xi _{t,r}u_{t-r} \\
&=&y_{t,nl}^{\hom }+y_{t,nl}^{par}\text{.}
\end{eqnarray*}%
as required.
\end{proof}

\section{APPENDIX}

\emph{TIME INVARIANT VECTOR FORM}

For the benefit of the reader this Section reviews some results on PARMA
models. \ Recall that the autoregressive coefficients are periodically
varying: $\phi _{m}(t)=\phi _{m}(\tau _{n})$ where $\tau _{n}=t-nl$. Recall
also that $t_{s}$ denotes time at the $s$th season: $t_{s}=Tl+s$, $%
s=1,\ldots ,l$, which written as $t_{s}-s=Tl$ is equivalent to $t_{s}\equiv
s $ $\func{mod}$ $l$. That is $t_{s}$ and $s$ are congruent modulo $l$ ($%
t_{s}$ and $s$ have the same remainder when they are divided by $l$).%
\footnote{%
The congruence class of $s$ modulo $l$ is given by%
\begin{equation*}
\lbrack s]_{l}=\{t_{s}\in \mathbb{Z}:t_{s}-s=Tl\text{, for some }T\in 
\mathbb{Z}\},
\end{equation*}%
and $[t_{s}]_{l}=[s]_{l}\Longleftrightarrow t_{s}\equiv s$ $\func{mod}$ $l$.
For example, if $l=4$, then there are four congruent classes which partition
the set $\mathbb{Z}$ into four disjoint sets:%
\begin{eqnarray*}
\lbrack 1]_{4} &=&\{ \pm 1,\pm 5,\pm 9,...\} \text{; }[2]_{4}=\{ \pm 2,\pm
6,\pm 10,...\}, \\
\lbrack 3]_{4} &=&\{ \pm 3,\pm 7,\pm 11,...\} \text{; }[4]_{4}=\{0,\pm 4,\pm
18,\pm 12,...\}.
\end{eqnarray*}%
} Thus, we can write $\phi _{m,s}\triangleq \phi _{m}(Tl+s)$ since $%
Tl+s\equiv s$ $\func{mod}$ $l$ (see eq. (\ref{PAR(P)})). \ We can see one of
the advantages of the elaborate notation that we employ in place of the
single index $t$, namely it conveys the point that the data generating
process of a time series variable depends on the season.

We assume without loss of generality that time $t$ is at the $l$th season,
that is $s=l$ (e.g., $t=t_{l}=(T+1)l$). Thus our $\mathbf{\Phi }_{\mathfrak{t%
},l}$ matrix in eq. (\ref{FAIM1}) will be denoted by $\mathbf{\Phi }(l)$ and
becomes:

\begin{equation*}
\mathbf{\Phi }(l)=\left( 
\begin{array}{cccccccc}
\phi _{1,1} & -1 &  &  &  &  &  &  \\ 
\phi _{2,2} & \phi _{1,2} & \ddots &  &  &  &  &  \\ 
\vdots & \vdots & \ddots & \ddots &  &  &  &  \\ 
\phi _{p,p} & \phi _{p-1,p} & \ddots & \ddots & \ddots &  &  &  \\ 
& \phi _{p,p+1} & \ddots & \ddots & \ddots & \ddots &  &  \\ 
&  & \ddots & \ddots & \ddots & \ddots & \ddots &  \\ 
&  &  & \phi _{p,l-1} & \phi _{p-1,l-1} & \cdots & \phi _{1,l-1} & -1 \\ 
&  &  &  & \phi _{p,l} & \cdots & \phi _{2,l} & \phi _{1,l}%
\end{array}%
\right)
\end{equation*}%
since for $m=1,\ldots ,\min (p,l)$, $r=0,\ldots ,l-m$, $\phi _{m}(t-r)=\phi
_{m,l-r}$ in eq. (\ref{FAIM1}). \ A convenient representation of the PAR
model in eq. (\ref{TVAR(P)}) is the VAR representation- hereafter we will
refer to it as the vector of seasons (VS) representation (see, for example,
Tiao and Guttman, 1980; Vecchia, 1985; Osborn, 1991; \ Franses, 1994,
1996a,b; Lund and Basawa, 2000; del Barrio Castro and Osborn, 2008).

The corresponding VS representation of the PAR($p;l$) model (ignoring the
drifts) is given by%
\begin{equation}
\mathbf{\Phi }_{0}\mathbf{y}_{T}=\mathbf{\Phi }_{1}\mathbf{y}_{T-1}+\cdots +%
\mathbf{\Phi }_{P}\mathbf{y}_{T-P}+\mathbf{\varepsilon }_{T},  \tag{B.1}
\end{equation}%
with $\mathbf{y}_{T}=(y_{1T},\ldots ,y_{lT})^{\prime }$, $\mathbf{%
\varepsilon }_{T}=(\varepsilon _{1T},\ldots ,\varepsilon _{lT})^{\prime }$,
where the first subscript refers to the season ($s$) and the second one to
the period ($T$). Moreover, $\mathbf{\Phi }_{0}=[\phi
_{ij}^{(0)}]_{i,j=1,\ldots ,l}$ is an $l\times l$ parameter matrix whose ($%
i,j$) entry is:%
\begin{equation}
\left \{ 
\begin{array}{lll}
1 & \text{if} & i=j, \\ 
0 & \text{if} & j>i, \\ 
-\phi _{i-j,i} & if & j<i,%
\end{array}%
\right.  \tag{B.2}
\end{equation}%
and $\mathbf{\Phi }_{1},\ldots ,\mathbf{\Phi }_{P}$ are $l\times l$
parameter matrices with ($i,j$) elements $\phi _{ij}^{(M)}=\phi _{i+lM-j,i}$%
, for $M=1,\ldots ,P$ (see for example Vecchia, 1985, Lund and Basawa, 2000,
Franses and Paap, 2005). The $l$-variate AR order $P$ is $P=[p/l]$, where $%
[x]$ denotes the smallest integer greater than or equal to $x$.

As pointed out by Franses (1994), the idea of stacking has been introduced
by Gladyshev (1961) and is also considered in e.g., Pagano (1978), Tiao and
Guttman (1980), Vecchia (1985), Osborn (1991), Franses (1994) and Lund and
Basawa (2000), who used it in the AR setting. The dynamic system in eq.
(B.1) can be written in a compact form%
\begin{equation*}
\mathbf{\Phi }(B)\mathbf{y}_{T}\mathbf{=\varepsilon }_{T}\text{ or }\left
\vert \mathbf{\Phi }(B)\right \vert \mathbf{y}_{T}\mathbf{=}adj[\mathbf{\Phi 
}(B)]\mathbf{\varepsilon }_{T}
\end{equation*}%
($adj(\mathbf{X})$ stands for the adjoint of matrix $\mathbf{X}$), where $%
\mathbf{\Phi }(B)=\mathbf{\Phi }_{0}-\Sigma _{M=1}^{P}\mathbf{\Phi }%
_{M}B^{M} $. Stationarity of $\mathbf{y}_{T}$ requires the roots of $%
\left
\vert \mathbf{\Phi }(z^{-1})\right \vert =0$ to lie strictly inside
the unit circle (see, among others, Tiao and Guttman, 1980; Osborn, 1991;
Franses, 1994, 1996a; Franses and Paap, 2005; del Barrio Castro and Osborn,
2008). For the ARMA($p,q;l$) model we replace $\mathbf{\varepsilon }_{T}$
with $\mathbf{u}_{T}=\mathbf{\Theta }(B)\mathbf{\varepsilon }_{T}$, where $%
\mathbf{\Theta }(B)=\mathbf{\Theta }_{0}-\Sigma _{N=1}^{Q}\mathbf{\Theta }%
_{N}B^{N}$ (see Lund and Basawa, 2000). The $l$-variate MA order $Q$ is $%
Q=[q/l]$. The moving average $l\times l$ parameter matrices $\{ \mathbf{%
\Theta }_{N}=[\theta _{ij}^{(N)}],$ $0\leq N\leq Q\}$ are obtained in a
similar manner to the AR matrices $\mathbf{\Phi }_{M}$ with $\theta
_{ij}^{(N)}$ replacing each occurrence of $\phi _{ij}^{(M)}$ (see Lund and
Basawa, 2000).

As an example, consider the PAR($2;4$) model 
\begin{equation*}
y_{t_{s}}=\phi _{1,s}y_{t_{s}-1}+\phi _{2,s}y_{t_{s}-2}+\varepsilon _{t_{s}},
\end{equation*}%
which can be written as%
\begin{equation*}
\mathbf{\Phi }_{0}\mathbf{y}_{T}=\mathbf{\Phi }_{1}\mathbf{y}_{T-1}+\mathbf{%
\varepsilon }_{T},
\end{equation*}%
for which the characteristic equation is%
\begin{equation*}
\left \vert \mathbf{\Phi }_{0}-\mathbf{\Phi }_{1}z\right \vert =\left \vert 
\begin{array}{llll}
1 & 0 & -\phi _{2,1}z & -\phi _{1,1}z \\ 
-\phi _{1,2} & 1 & 0 & -\phi _{2,2}z \\ 
-\phi _{2,3} & -\phi _{1,3} & 1 & 0 \\ 
0 & -\phi _{2,4} & -\phi _{1,4} & 1%
\end{array}%
\right \vert =0.
\end{equation*}%
Hence, when the nonlinear parameter restriction 
\begin{eqnarray*}
&&\left \vert \phi _{2,2}\phi _{1,3}\phi _{1,4}+\phi _{2,2}\phi _{2,4}+\phi
_{2,1}\phi _{1,2}\phi _{1,3}+\phi _{2,1}\phi _{2,3}+\phi _{1,1}\phi
_{1,2}\phi _{1,3}\phi _{1,4}\right. \\
&&\left. +\phi _{1,1}\phi _{1,2}\phi _{2,4}+\phi _{1,1}\phi _{1,4}\phi
_{2,3}-\phi _{2,1}\phi _{2,2}\phi _{2,3}\phi _{2,4}\right \vert <1,
\end{eqnarray*}%
is imposed on the parameters, the VS representation of the PAR($2;4$) model
is stationary (see Franses and Paap, 2005). When $\phi _{2,s}=0$ for all $s$%
, that is we have the PAR($1;4$) model, then the stationarity condition
reduces to: $\left \vert \phi _{1,1}\phi _{1,2}\phi _{1,3}\phi
_{1,4}\right
\vert <1$ which is equivalent to our condition $\left \vert
\xi _{t,l}\right
\vert <1$ or, to put it in another way, the absolute value
of $\left \vert \mathbf{\Phi }(l)\right \vert $ is less than one.

\end{document}